\documentclass[runningheads]{llncs}

\usepackage[T1]{fontenc}
\usepackage{algorithm}

\usepackage{amsmath}
\usepackage{booktabs}
\usepackage{algpseudocode}
\usepackage{amsmath}
\usepackage{fontawesome}
\usepackage{multirow}
\usepackage{graphicx}
\usepackage{nicefrac}
\usepackage{adjustbox}
\usepackage{graphicx}
\usepackage{soul}
\usepackage{xcolor}

\sethlcolor{yellow} 
\usepackage{subcaption}
\usepackage[colorlinks=true]{hyperref}
\usepackage{comment}
\usepackage{amssymb,amsmath}
\usepackage{mathtools}
\usepackage{stmaryrd}

\usepackage{tabularx} 
\usepackage{xfp}
\usepackage{tikz}
\usepackage{tikzscale}
\usepackage{xspace}
\usepackage{pgfplots}
\usepgfplotslibrary{groupplots}
\pgfplotsset{compat=1.16}
\usepackage{xparse}
\usepackage{graphicx}
\usepackage{orcidlink}

\begin{document}

\title{Enhancing Privacy Preservation and Reducing Analysis Time with Federated Transfer Learning in Digital Twins-based Computed Tomography Scan Analysis}

\titlerunning{FTL for Digital Twins-based CT Scan Analysis}

\author{Avais Jan \inst{1}\orcidlink{0009-0006-0199-0984} \and
Qasim Zia\inst{1}\orcidlink{0009-0004-2028-5960} \and
Murray Patterson\inst{1}\orcidlink{0000-0002-4329-0234}}

\authorrunning{Avais Jan et al.}

\institute{Georgia State University, Atlanta, GA, 30303 USA \\
\email{\{ajan3,qzia1\}@student.gsu.edu, mpatterson30@gsu.edu} 
}

\maketitle              

\begin{abstract}
The application of Digital Twin (DT) technology and Federated Learning (FL) has great potential to change the field of biomedical image analysis, particularly for Computed Tomography (CT) scans. This paper presents Federated Transfer Learning (FTL) as a new Digital Twin-based CT scan analysis paradigm. FTL uses pre-trained models and knowledge transfer between peer nodes to solve problems such as data privacy, limited computing resources, and data heterogeneity. The proposed framework allows real-time collaboration between cloud servers and Digital Twin-enabled CT scanners while protecting patient identity.

We apply the FTL method to a heterogeneous CT scan dataset and assess model performance using convergence time, model accuracy, precision, recall, F1 score, and confusion matrix. It has been shown to perform better than conventional FL and Clustered Federated Learning (CFL) methods with better precision, accuracy, recall, and F1-score. The technique is beneficial in settings where the data is not independently and identically distributed (non-IID), and it offers reliable, efficient, and secure solutions for medical diagnosis.
These findings highlight the possibility of using FTL to improve decision-making in digital twin-based CT scan analysis, secure and efficient medical image analysis, promote privacy, and open new possibilities for applying precision medicine and smart healthcare systems. 
\keywords{Digital Twin, Federated Learning, Federated Transfer Learning, Computed Tomography (CT) Scan, Privacy Preservation}
\end{abstract}

\section{Introduction}
Medical imaging is one of the most essential parts of the modern healthcare system and helps doctors make the right decisions regarding the treatment of the patient and his or her further therapy. Among all modalities, computed tomography (CT) scans produce cross-sectional images that help diagnose diseases, fractures, and infections \cite{hussain2022modern}. However, the centralized nature of traditional CT scan interpretation poses several challenges, such as data privacy, computational cost, and lack of collaboration between different institutions \cite{shakor2024recent}. 

Federated Learning (FL) is a form of machine learning where many clients, for example hospitals, medical institutions, or edge devices, can train a common model without exchanging their data locally. Federated learning has emerged as an effective technique to overcome challenges such as these through the training of multiple medical institutions in machine learning models without exchanging patient raw data \cite{nguyen2022federated}. This decentralized framework improves privacy by keeping data in the client's digital profile and yet improves collective intelligence \cite{cirillo2025applications}. However, conventional FL approaches are sensitive to domain shifts between different institutions, resulting in performance deterioration. To this end, Federated Transfer Learning (FTL) extends the FL framework with transfer learning techniques to enable knowledge transfer from pre-trained models and then fine-tunes them for the given local data distributions \cite{dai2023addressing}.

A Digital Twin is a real-time virtual copy of a system, process, or entity that continuously receives data from its physical counterpart.
Digital twins (DT) are becoming popular for monitoring and proactive care in healthcare. The digital twin of a patient's CT scan can model the disease evolution, support individualized care, and shape decision-making \cite{subasi2024digital}. Thus, the application of FTL to DT-based CT scan analysis makes it possible to improve the generalization of models, respect data privacy, and save computing resources. 
 
This research designs an FTL-based digital twin framework for CT scan analysis to balance privacy, time, and accuracy. The main contributions of this work are:
\begin{enumerate}
    \item Proposing an FTL framework for CT scan analysis in DT environments for private and efficient model training.
    \item Optimization techniques to improve analysis speed without sacrificing diagnosis level.
    \item Comparisons between FTL, Clustered FL, and conventional FL highlight FTL's accuracy and performance benefits.  
\end{enumerate}

This paper is structured as follows: related works are described in Section \ref{sec:related_work}. We suggest an architecture of an FTL-based DT-CT scan model in Section \ref{sec:ftl_based_dtscan}. The comprehensive experiments are then carried out and assessed in Section \ref{sec: eval} to determine the effectiveness and efficiency of our suggested architecture. We finally arrive at our conclusions in Section \ref{sec:concl}.

\section{Related Works}\label{sec:related_work}
Federated Learning(FL) is a well-known model for privacy-preserving decentralized training without the exchange of sensitive data. Various studies have been conducted on FL in medical imaging, especially for analyzing computed tomography (CT) scans. For example, FL's feasibility was shown by Sheller et al. \cite{sheller2019multi} in brain tumor segmentation while preserving data privacy to the level of central training. Using a priority algorithm, Zia et al.\cite{zia2016improving} improve response time.  Likewise, Dou et al.\cite{dou2021federated} investigated FL for the detection of lung disease on CT scans and pointed out its benefits in data security and collaborative learning in hospitals.  Zia et al. \cite{zia2025optimized} discussed optimized decision-making with the help of EfficientNet in DT-VANET.

However, conventional FL experiences challenges in domain change when deployed in heterogeneous medical datasets. To resolve this, Federated Transfer Learning (FTL) has become an effective approach to expand on pre-trained models and then fine-tune them for target domains. Yin et al.\cite{yin2024federated} introduced an FTL framework for cross-institutional medical image analysis, enhancing the generalization capacity of the model on unseen data sets. Zia et al.\cite{zia2024priority} improve communication using a priority algorithm in a network based on Digital Twin (DT). However, Irfan et al. \cite{irfan2024federated} have integrated FTL with an attention mechanism to improve feature extraction in chest CT scan classification, thus reducing misclassification rates. Zia et al. \cite{zia2025hierarchical} have discussed the use of FTL in DT-VANET.

However, several challenges in such FTL approaches related to computational efficiency and convergence time are inherent. These challenges arise from the high communication overhead during model aggregation, the heterogeneous data distributions across clients that slow convergence, and the varying computational resources at different nodes. Zhang et al.\cite{zhang2021optimizing} had also proposed optimized model aggregation techniques to reduce communication overhead in FL, but their application to real-time medical imaging has been limited. Zia et al.\cite{zia2015survey} discuss different protocols necessary for network communication. Furthermore, recent work in digital twin (DT)--based healthcare systems, including those of Saha et al. \cite{sahal2022personal}, has explored DT models for personalized patient monitoring. Therefore, integrating FTL into DT frameworks offers a promising direction to enhance privacy and computational efficiency in medical imaging. A DT provides a continuous virtual model of the physical imaging system and continuous real-time data synchronization and predictive analytics. This integration leads to more adaptive model updates and more efficient resource allocation that can help alleviate the computational burdens and convergence delays typical of traditional FL systems. Consequently, by taking advantage of the real-time and personalization of DT monitoring functionality, the overall framework can improve privacy and reduce the analysis time in CT scan analysis.

Based on these previous works, our research aims to optimize real-time decision-making by combining DT with FTL in the context of CT scan analysis. Our approach also addresses privacy issues and reduces the analysis time, to help overcome the accuracy-feasibility gap in federated medical imaging. 
\begin{figure*}[!ht]
    \centering
    \includegraphics[width=\textwidth]{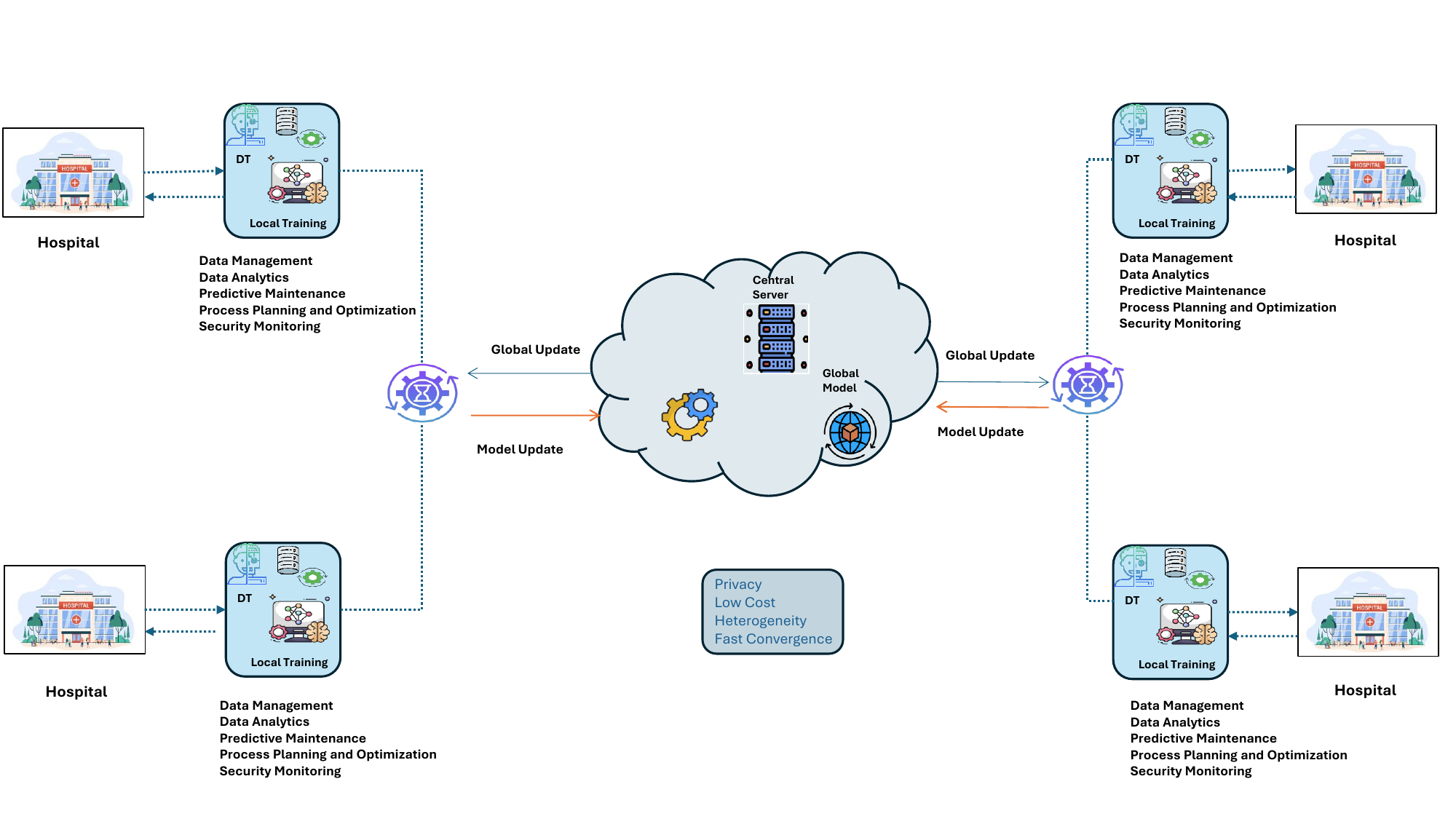}
    \caption{High-level architecture of Federated Transfer Learning for digital twin-based Computed Tomography scan.}
    \label{fig_1}
\end{figure*}

\section{Proposed Approach} \label{sec:ftl_based_dtscan}
This study presents a Federated Transfer Learning (FTL) framework to support the analysis of Digital Twin-based Computed Tomography (DT-Scan). The proposed model is intended to preserve privacy while significantly reducing the analysis time for medical imaging applications. In contrast to conventional centralized approaches, FTL enables several medical healthcare centers to jointly train deep learning models without exchanging raw patient data to comply with data privacy regulations such as HIPAA and GDPR.

\subsection{FTL Method Architecture } 
This section provides a high-level architecture for FTL in DT-Scan, as seen in Figure \ref{fig_1}. The proposed DT-based FTL method consists of three main components:
\subsubsection{Medical Institutions (Clients)}
Every participating hospital and diagnostic center maintains its local CT image data set. A local model such as DenseNet-121 is deployed and trained on local patient data. Raw images are not shared, as gradients representing local model updates are exchanged between institutions.
\subsubsection{Federated Server (Global Aggregator)}
The server gathers model updates from various institutions and then applies Federated Averaging (FedAvg). The aggregator uses transfer learning methods to join data sources while accommodating differences in CT scan features. The globally updated model is returned to local nodes as a reference for additional refinement. 
\subsubsection{Digital Twin for CT-Scan Analysis}
A virtual digital twin monitors the progression of the disease by assigning each patient's CT scan to this model. The FTL model provides real-time analysis with optimized inference speed for diagnosis. Digital Twin models are improved through continuous federated updates, which leads to growing accuracy over time.

This architecture has a few advantages. First, since raw medical images are not exchanged, we have a reduced risk of data breaches. Second, because of its transfer learning capabilities, DenseNet-121 accelerates model convergence during analysis and reduces analysis time. Furthermore, improved model generalization improves diagnostic accuracy by aggregating knowledge from various sources. Lastly, scalability and adaptability can extend the FTL framework to multiple healthcare institutions with varying data distributions.

\subsection{Cloud Server Federated Transfer Learning }
This algorithm represents a federated transfer learning process in which the central server updates its global model parameters \( W_{\text{global}} \) by collaborating with multiple hospitals (CT scanners in our case). First, the model parameters of the federated transfer learning process are already on the central cloud server. The group of hospitals \( \mathcal H_n \) that have not yet taken part in Federated Transfer Learning is initialized. The process repeats several update rounds. The global model parameters \( W_{\text{global}} \) are sent to a Digital Twin of a selected hospital in each round. Using the parameters received, denoted by \( l = \text{LOCAL-UPDATE}(W_{\text{global}^-}) \), the hospital then performs a local update. A weighting function \( F_{weight} \) (\(W_{\text{global}^+}\)) is then used to update the global model which is explained further in detail next section. After the update, the selected hospital is removed from the list of participating hospitals since every hospital should participate in the training process. This is done until all updates are made and the optimized global model parameters \( W_{\text{global}} \) are acquired. 
\begin{algorithm}
\caption{\scriptsize Weighted Cloud Server Cycling Model Update}
\begin{scriptsize} 
\begin{algorithmic}[1]
\State The central server has existing model parameters \( W_{\text{global}} \);
\State Initialize \( \mathcal H_n \) the set of participating hospitals;
\For{each update \( j = 0, 1, 2, \ldots, j-1 \)}
    \State \(W_{\text{global}^-}\) = \( W_{\text{global}} \);
    \State Select hospital i from \( \mathcal {H}_n \);
    \State Send \(W_{\text{global}^-}\) to the Digital Twin of hospital i;
    \State \( l = \text{LOCAL-UPDATE}(W_{\text{global}^-}) \);
    \State Calculate \( F_{weight} \) (\(W_{\text{global}^+}\));
    \State Update $W_{global}$ ;
    \State Eliminate i from the remaining hospitals $H_n$;
\EndFor
\State Output the final global model parameters \( W_{\text{global}} \);
\end{algorithmic}
\end{scriptsize}
\end{algorithm}
\subsection{Hospital Local Update Federated Transfer Learning }
This algorithm describes the hospital's local update process in Federated Transfer Learning. This algorithm aims to fine-tune the parameters of the global model \(W_{\text{global}^-}\) at the hospital level before sending them back to the central server. The process is divided into two main parts and briefed as follows: the operations carried out by the hospital server and its digital twin(CT scanner in our case). The hospital server receives the global model parameters \(W_{\text{global}^-}\) from the central server and then forwards it to its local digital twin system. The digital twin (of CT scanner in our case) then fine-tunes the model using its local dataset $D_i$ and produces the updated model parameters \(W_{\text{global}^+}\). The fine-tuned parameters are then forwarded to the hospital server, which sends them to the central server. This process helps to improve the global model with hospital-level information before the model makes its contribution to the federated transfer learning system. 
\newcommand{\Input}[1]{\item[\textbf{Input}] #1}
\newcommand{\Output}[1]{\item[\textbf{Output}] #1}
\begin{algorithm}
\caption{\scriptsize Hospital Local Update Federated Transfer Learning}
\begin{scriptsize} 
\begin{algorithmic}[1]
\Input: The model parameters of the hospital that need to be fine-tuned  \(W_{\text{global}^-}\);
\Output: The fine-tuned model parameters \(W_{\text{global}^+}\);
\Procedure{Local-Update}{}
    \State \textbf{(I) For the hospital server:}
    \State Receive \(W_{\text{global}^-}\) from the central server;
    \State Send \(W_{\text{global}^-}\) to the local digital twin system;
    \State Receive \(W_{\text{global}^+}\) from the local digital twin system;
    \State Send \(W_{\text{global}^+}\) to the central server;
    \State \textbf{(II) For the hospital's digital twin:}
    \State Receive \(W_{\text{global}^-}\) from the hospital server;
    \State Fine-tune the model:
    \[
        W_{\text{global}^+} = \text{FINE-TUNE}( W_{\text{global}^-}, D_i);
    \]
    \State Send \(W_{\text{global}^+}\) to the hospital server;
\EndProcedure
\end{algorithmic}
\end{scriptsize}
\end{algorithm}

\section{Performance Evaluation}\label{sec: eval}
This section uses various performance evaluations to assess our proposed framework using Federated Transfer Learning.
\subsection{Dataset}
\begin{figure*}[!ht]
    \centering
    \begin{tabular}{cc}
        \includegraphics[width=0.40\textwidth]{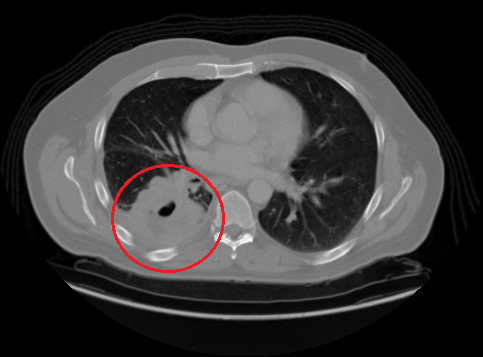} &
        \includegraphics[width=0.37\textwidth]{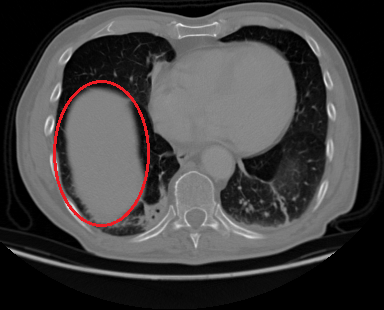} \\
        \textbf{Adenocarcinoma} & \textbf{Large Cell Carcinoma} \\
        \includegraphics[width=0.40\textwidth]{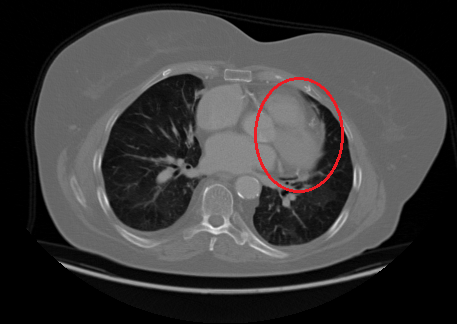} &
        \includegraphics[width=0.37\textwidth]{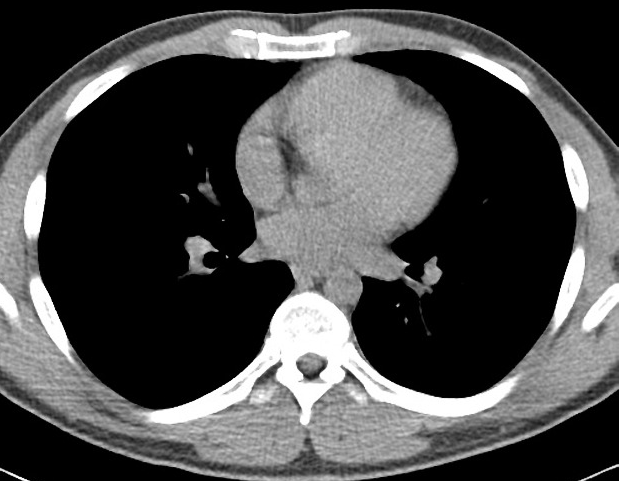} \\
        \textbf{Squamous Cell Carcinoma} & \textbf{Normal CT-Scan} \\
    \end{tabular}
    \caption{Sample of computed tomography scan images from the dataset ~\cite{hany2020chest} }
    \label{fig_2}
\end{figure*}

We used the Chest CT-Scan images\footnote{\scriptsize\url{https://www.kaggle.com/datasets/mohamedhanyyy/chest-ctscan-images?datasetId=839140}} \cite{hany2020chest} dataset for our simulations. This is one of the most commonly used datasets for CT scan images. Three distinct forms of chest cancer (adenocarcinoma, large cell carcinoma, squamous cell carcinoma), as well as normal CT-Scan images, are included in this dataset, as mentioned in Figure \ref{fig_2}, in which the red circles indicate the affected regions. These CT scan images reveal differences in cancer types, imaging protocols, resolution, and contrast across different hospitals while still being useful to build a single, robust global model. Through these various cancer types appearing in CT scans, the visuals further highlight the problem of accurate tumor detection and classification. 
The features of the dataset are mentioned in the Table \ref{tab:table1}. 

\begin{table}[h!]
\centering
\caption{Features of the Chest Computed Tomography Scan Dataset}
\label{tab:table1} 
\resizebox{0.9\textwidth}{!}{  
\begin{tabularx}{\textwidth}{@{}>{\raggedright\arraybackslash}X >{\raggedright\arraybackslash}X >{\raggedright\arraybackslash}X >{\raggedright\arraybackslash}X@{}}
\hline
\textbf{Cancer Types} & \textbf{Origin} & \textbf{Grade/Severity} & \textbf{Symptoms} \\
\hline\hline
Adenocarcinoma & Outer regions of lungs & Varies & Persistent cough, hoarseness, weight loss, weakness \\
Large Cell Carcinoma & Anywhere in lungs & High (aggressive) & Rapid progression, large masses, necrosis \\
Squamous Cell Carcinoma & Central lung, near bronchi & Varies & Cough, airway obstruction, cavitation, weight loss \\
\hline
\end{tabularx}
}
\end{table}

\subsection{Baseline Studies}
We will compare techniques such as FL, CFL, and FTL. We chose FL and CFL for comparison because they are closely related to FTL. We have used the DenseNet-121 model because it is pre-trained on the medical dataset CheXpert. So, there is no need to simulate data partition among hospitals. As our analysis is focused on how these three approaches perform and the convergence time when have data at the client(hospital) side. So, we are also not considering any latency issues caused during communication between the cloud and hospital server.
\subsection{Performance Metrics}
The efficiency of the suggested FTL framework for Digital Twins-based CT scan analysis will be assessed using various critical performance criteria.
\subsubsection{Accuracy}
Accuracy represents the proportion of correct classifications of CT scan images out of the total samples used in the model. It is computed as:
\begin{equation}
\text{Accuracy} = \frac{\text{TP} + \text{TN}}{\text{TP} + \text{TN} + \text{FP} + \text{FN}}
\end{equation}
TN, TP, FN, and FP represent true negatives, true positives, false negatives,  and false positives, respectively.

\subsubsection{Precision}
Precision calculates the percentage of actual positive instances correctly identified among all the cases predicted as positive. It is defined as:
\begin{equation}
\text{Precision} = \frac{\text{TP}}{\text{TP} + \text{FP}}
\end{equation}
A higher precision represents fewer false positives.
\subsubsection{Recall (Sensitivity)}
Recall measures how many actual positives were correctly classified. It is calculated as:
\begin{equation}
\text{Recall} = \frac{\text{TP}}{\text{TP} + \text{FN}}
\end{equation}
The high recall value shows the model's ability to effectively identify positive cases.
\subsubsection{F1-Score}
We need a metric that can capture precision and recall to understand how well a model performs in classification comprehensively. This is where the F1 score comes in. It is calculated as:
\begin{equation}
\text{F1-Score} =  \frac{2 \times\text{Precision} \times \text{Recall}}{\text{Precision} + \text{Recall}}
\end{equation}
The model achieved a higher F1 score, providing optimal precision and recall trade-off.
\begin{equation}
\text{Overall Accuracy} = \frac{\sum \text{True Positives}}{\text{Total Samples}} = \frac{\sum_{i=1}^{n} \text{TP}_i}{\sum_{i=1}^{n} \text{Total Samples}_i}
\end{equation}
Where: \(\text{True Positives}_i\) is the number of correct predictions of class \( i \) (diagonal elements of the confusion matrix),

\( \text{Total Samples}_i \) is the total amount of class \( i \) samples (sum of each row in the confusion matrix),

\( n \) is the number of classes.

The convergence time is known as the time when the FTL model converges to an optimal value during training. This metric allows us to compare the efficiency of the proposed approach with federated learning and clustered federated learning paradigms. The model is considered to have converged when any improvement in accuracy is less than a predetermined threshold and when the change in loss between consecutive rounds is less than a minimal threshold for several rounds in sequence.

The confusion matrix is also calculated based on TP, TN, FP, and FN. These metrics will allow the study to show that Federated Transfer Learning improves privacy, reduces analysis time, and still provides high classification performance in CT scan analysis.

\subsection{Experimental and Performance Analysis }
This section will discuss the performance of the proposed FTL approach in Digital Twins-based CT scan analysis. To this end, we design experiments to assess the time taken for convergence and the model's accuracy. We will compare our FTL-based method with traditional FL and CFL approaches. 

\begin{figure}[h]
    \centering
    \includegraphics[width=0.328\linewidth]{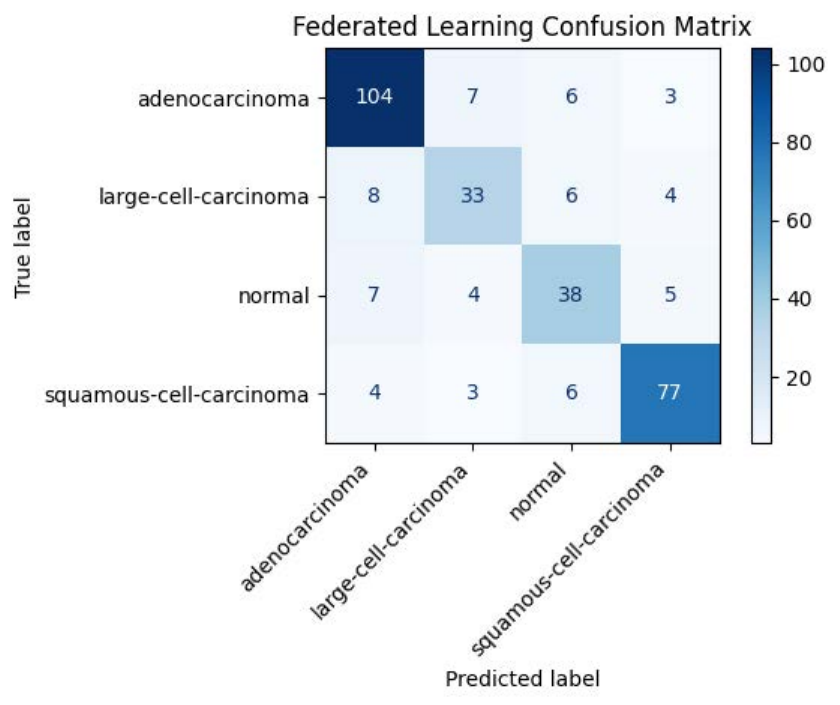} 
    \includegraphics[width=0.328\linewidth]{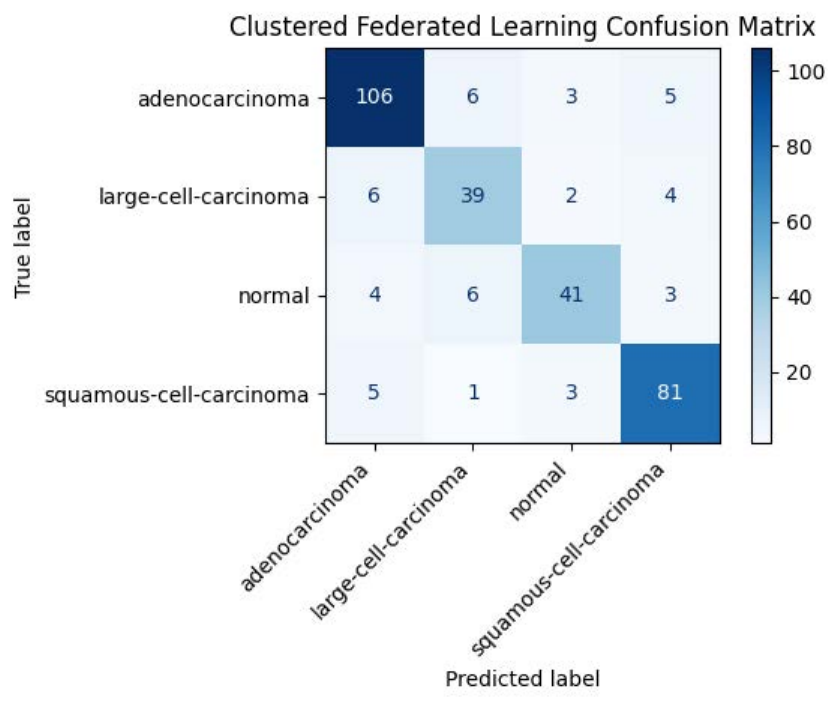}
    \includegraphics[width=0.328\linewidth]{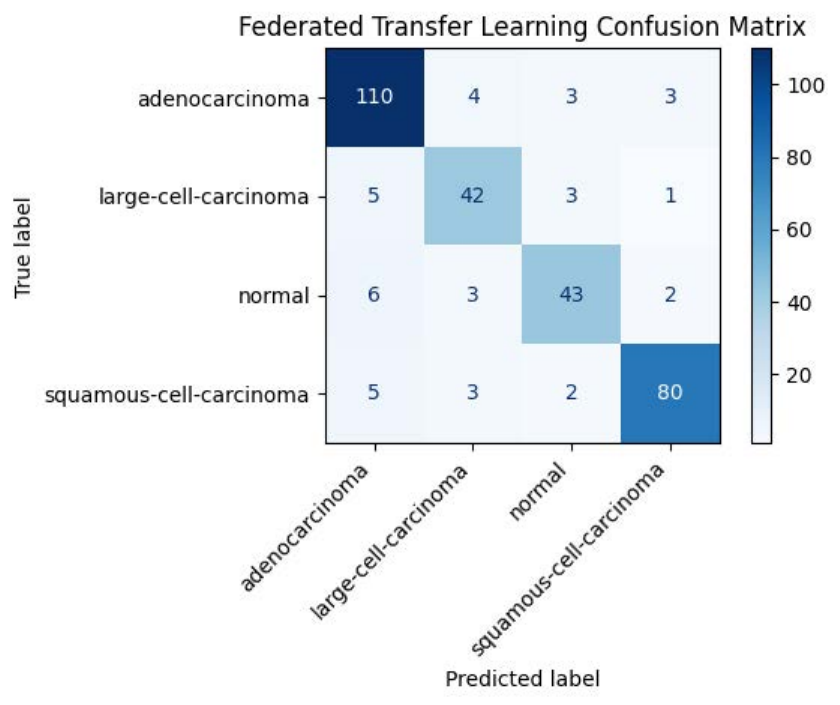} \\
    \caption{Comparison of confusion matrices for Federated Learning, Clustered Federated Learning, and Federated Transfer Learning}
    \label{fig:confusion_matrices}
\end{figure}

FTL's performance is compared using various performance metrics that are very important for reducing analysis time and improving privacy, which includes test accuracy, recall, F1 score, and precision in Table~\ref{tab:combined_metrics}. The overall accuracy of Federated Learning is 0.8000, clustered Federated Learning is 0.8476, and Federated Transfer Learning is 0.8730. The evaluation of these methods for different cancer types demonstrates robust performance, particularly for Adenocarcinoma and Squamous Cell Carcinoma, indicating that these models effectively distinguish between different cancer types. Notably, FTL exhibits the highest accuracy and F1-score for Adenocarcinoma and Squamous Cell Carcinoma, suggesting that transfer learning enhances the model's ability to generalize from pre-trained knowledge, leading to improved classification performance. This improvement is likely due to the leveraging of pre-trained models, which provide a strong foundation for learning, thereby reducing the convergence time and enhancing efficiency. The classification metrics are assessed using the confusion matrix as shown in Figure \ref{fig:confusion_matrices}.

However, the performance varies for Large Cell Carcinoma and Normal classes, where the precision and recall are relatively lower compared to the other classes. This discrepancy may be attributed to the smaller sample size for these classes, which can affect the model's ability to learn distinctive features. Despite this, CFL shows promise in improving the recall for Large Cell Carcinoma, indicating that clustering similar data points can help better capture the underlying patterns. Overall, the results demonstrate the potential of federated learning techniques in medical imaging, particularly in scenarios where data privacy is paramount. The convergence behavior, as illustrated in the figures, further supports the efficiency of FTL, which outperforms standard FL and CFL in terms of training time, making it a viable approach for real-world applications.

\begin{figure}[h]
    \centering
    \includegraphics[width=0.832\textwidth]{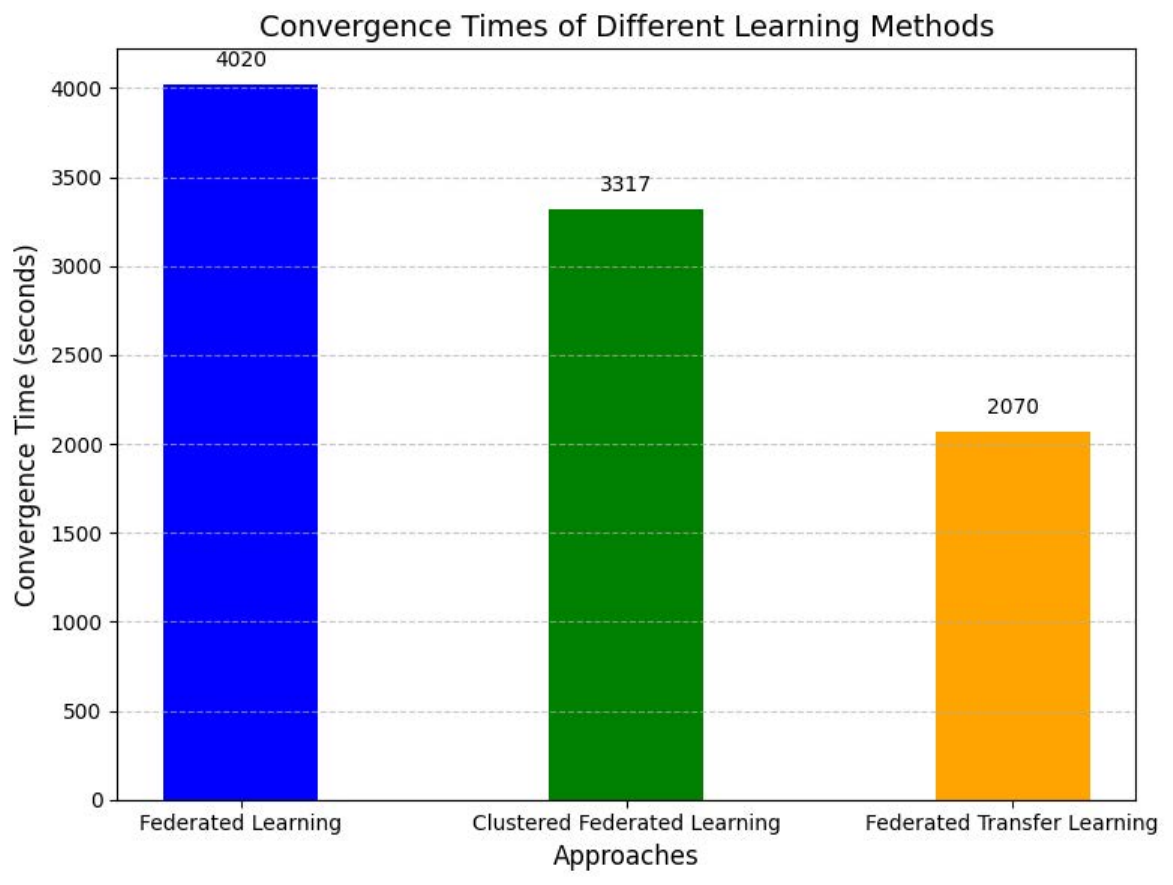} 
    \caption{Comparison of convergence time for Federated Learning, Centralized Federated Learning, and Federated Transfer Learning for classification of different cancer types from computed tomography scans}
    \label{fig:CT}
\end{figure}

\begin{table}[h!]
\centering
\scriptsize
\begin{tabular}{lcccccc}
\hline
\textbf{Method} & \textbf{Classes} & \textbf{\# of Images} & \textbf{Precision} & \textbf{Recall} & \textbf{F1-Score} & \textbf{Accuracy} \\
\hline
& Adenocarcinoma & 120 & 0.8455 & 0.8667 & 0.8560 & 0.8667 \\
FL  & Large Cell Carcinoma & 51 & 0.7021 & 0.6471 & 0.6735 & 0.6471 \\
 & Normal & 54 & 0.6786 & 0.7037 & 0.6909 & 0.7037 \\
 & Squamous Cell Carcinoma & 90 & 0.8652 & 0.8556 & 0.8603 & 0.8556 \\
\hline
 & Adenocarcinoma & 120 & 0.8760 & 0.8833 & 0.8797 & 0.8833 \\
CFL & Large Cell Carcinoma & 51 & 0.7500 & 0.7647 & 0.7573 & 0.7647 \\
 & Normal & 54 & 0.8367 & 0.7593 & 0.7961 & 0.7593 \\
 & Squamous Cell Carcinoma & 90 & 0.8710 & 0.9000 & 0.8852 & 0.9000 \\
\hline
 & Adenocarcinoma & 120 & 0.8730 & 0.9167 & 0.8943 & 0.9167 \\
FTL & Large Cell Carcinoma & 51 & 0.8077 & 0.8235 & 0.8155 & 0.8235 \\
 & Normal & 54 & 0.8431 & 0.7963 & 0.8190 & 0.7963 \\
 & Squamous Cell Carcinoma & 90 & 0.9302 & 0.8889 & 0.9091 & 0.8889 \\
\hline
\end{tabular}
\caption{Comparison of evaluation metrics for lung cancer type classification using Federated Learning, Clustered Federated Learning, and Federated Transfer Learning
}
\label{tab:combined_metrics}
\end{table}

\section{Conclusion}\label{sec:concl}
This research paper aims to explore whether FTL can improve privacy protection and speed up the analysis in the application of Digital Twins-based Computed Tomography (CT) scan analysis. The results show that FTL can preserve patients' privacy by using decentralized learning on several local datasets to train the model and still achieve high accuracy in CT scan analysis. As a result, no sensitive data needs to be shared between institutions, thus fulfilling privacy policies like GDPR and HIPAA. 

Additionally, the application of transfer learning can help the model learn quickly and with less computational time than the traditional federated and clustered federated learning processes. Digital Twin technology and system integration improve real-time simulation and monitoring of CT scan processes, thus improving accuracy and efficiency.

Therefore, in conclusion, FTL can be regarded as a potential improvement in protecting privacy in medical imaging and reducing the time taken to analyze the images. Future work should include improving the size model and checking its effectiveness in clinical settings to confirm its efficacy and robustness. This research used real-time simulation by updating the digital model with current data at regular intervals from the CT scanners. The digital twin of the CT scanner is updated in real time to replicate the actual process, thus enabling the monitoring of the system for proactive maintenance. Digital Twin technology integrated into our Federated Transfer Learning framework enables online updates of the model based on real-time data. The results of the experiments show that the performance metrics and the model convergence are improved, which are the immediate advantages of having a system that can simulate and observe the CT scan processes online.

\end{document}